# *Meraculous-2D*: Haplotype-sensitive Assembly of Highly Heterozygous genomes.

Eugene Goltsman [1], Isaac Ho [1], Daniel Rokhsar [1,2,3]

[1] DOE Joint Genome Institute, 2800 Mitchell Drive, Walnut Creek, CA 94598

[2] Department of Molecular and Cell Biology, University of California, Berkeley CA 94720

[3] Molecular Genetics Unit, Okinawa Institute of Science and Technology, Onna, Japan

**Abstract**

Highly polymorphic diploid genomes are notoriously difficult to assemble to a high degree of contiguity and accuracy. While many short read assemblers attempt to simplify the de Brujin graph by identifying and resolving variant-induced bubbles to produce a haploid mosaic result, this approach is only viable when variants are relatively rare and the bubbles are well defined in a graph context. We observed that diploid genomes with very high levels of heterozygosity fail to display well-resolved bubble structures in a typical assembly graph and thus result in highly fragmented and incomplete assemblies when handled by the Meraculous2 assembler. Here we present an enhancement of Meraculous2 algorithm, called Meraculous-2D, which preserves haplotypes across variant sites and generates accurate assembly of highly heterozygous diploid genomes. Preserving and taking advantage of the allelic variation throughout the assembly process allows reconstructing both haplomes at once, without the need to pick arbitrary paths through bubble structures. In this approach, regions of homozygosity, and not variation, are viewed as the main source of ambiguity during scaffolding, since in these regions distinct haplotypes collapse. We resolve this ambiguity by duplicating collapsed homozygous segments that have strong but conflicting linkage to haplotype-specific contigs. We also enhanced the original diploidy resolution method of Meraculous2 to maintain and report *phased* haplotype variant information when a pair of local paths through a variant region can be confidently identified using read- or mate-pair phasing information. Meraculous-2D thus includes two diploid resolution modes – one a modification of the original method in which a single mosaic haplotype is reported, best used with diploid genomes with low heterozygosity, and a second novel method for highly heterozygous data sets.

Availability: Meraculous-2D is available under the GNU General Public License from https://sourceforge.net/projects/meraculous20/

# 1. INTRODUCTION

In the de Brujin graph approach, which is the most common method for assembling short-read datasets (reviewed in {Miller 2010}, {Simpson 2015}), the presence of allelic variants in a diploid genome means that even for regions that are free of repeats there is not a single unique path through the sub-graph.{Fasulo 2002} The problem is most acute for marine invertebrates (e.g., {Dehal et al. 2002}, {Vinson et al. 2005}, {Sea Urchin 2006}, {Putnam et al 2008}, {Zhang et al. 2012}) whose high population sizes lead to nucleotide heterozygosity above 1-2%, and in outbred or hybrid plant species where the progenitor genomes can be highly divergent from each other (*e.g*., {Wu et al. 2014}). In these scenarios, long contigs and scaffolds are challenging to produce, and various approaches are used to collapse homologous regions into a single 'consensus' sequence that is a mosaic of the two haplotypes. While such mosaic haplotypes may be useful as a basis for genome analysis and resequencing, they do not accurately capture the true sequence of a diploid genome, especially in the case of interspecific hybrids where the collapsed mosaic sequence represents neither haplome. High levels of heterozygosity may also lead to inaccurate representation of repetitive elements, since allelic variants and dispersed copies may be confounded.

Many assemblers, including previous versions of Meraculous,[Chapman 2016] attempt to reduce the complexity of the de Brujin graph by identifying and simplifying variant-induced structures. An isolated SNP, for example, will manifest itself as a *bubble* in the graph, which, once identified, can be traversed by selecting one of the two paths and omitting or masking the other [Kajitani 2014, Bankevich 2012, Zerbino 2008]. Longer regions of variation can be traversed in similar ways. A key parameter is the product of the nucleotide heterozygosity $h$ and the k-mer size $k$. When $hk$ is small compared to 1, heterozygous sites are rare, and bubbles are small and isolated. However, when $hk$ is large, the resulting bubbles typically span multiple variant sites and the alternate paths across the bubble increase in length, becoming more difficult to define and detect. This happens because a longer variant region is less likely to manifest itself as two simple paths that converge at both ends, which is a typical prerequisite for a bubble. Instead, the paths through it are more likely to be terminated or confounded by other factors such as repeats, physical gaps, sequencing errors, *etc*., and thus are not easily recognized in the de Bruijn graph without bringing additional information (*e.g*., paired-ends) into play.

In Meraculous2, a UUtig (unique path through the de Bruin graph) must terminate when the next k-mer extension candidate is either an X- or and F-type k-mer, meaning, a physical coverage gap or a fork follows (for nomenclature, see {Chapman 2016}). When the nucleotide heterozygosity is low with respect to the chosen kmer size ($h \ll 1/k$), we can expect variant-containing contigs to be short (length 2k-1 for an isolated SNP) and to converge on unique kmer extensions at both ends -- a clear bubble scenario. But with the nucleotide heterozygosity increasing above $1/k$, the bubble detection rate falls sharply, as seen in Figure 1. (Heterozygosity in this context should always be envisioned relative to the chosen k-mer size since in the same variant region, shorter k-mers will result in higher bubble resolution.) Meraculous2 dealt with uncaught bubbles by filtering out all remaining *non-*bubble contigs (a.k.a. *isotigs*) that displayed a depth profile in line with the *half-depth* peak of the bi-modal contig k-mer depth distribution. This works well for low-heterozygosity datasets ($hk \ll 1$) since most gaps created by such filtering are typically small and therefore do not disrupt scaffolding, and can be easily filled during the gap closure stage. In a highly heterozygous scenario, however, this is not an acceptable solution as it can potentially eliminate a very large fraction of the genome prior to the scaffolding step. To account for this scenario, which is becoming more and more common with increasing read lengths and quality and, therefore, higher optimal k, we developed an additional diploid assembly mode that takes advantage of haplotype divergence instead of masking it.

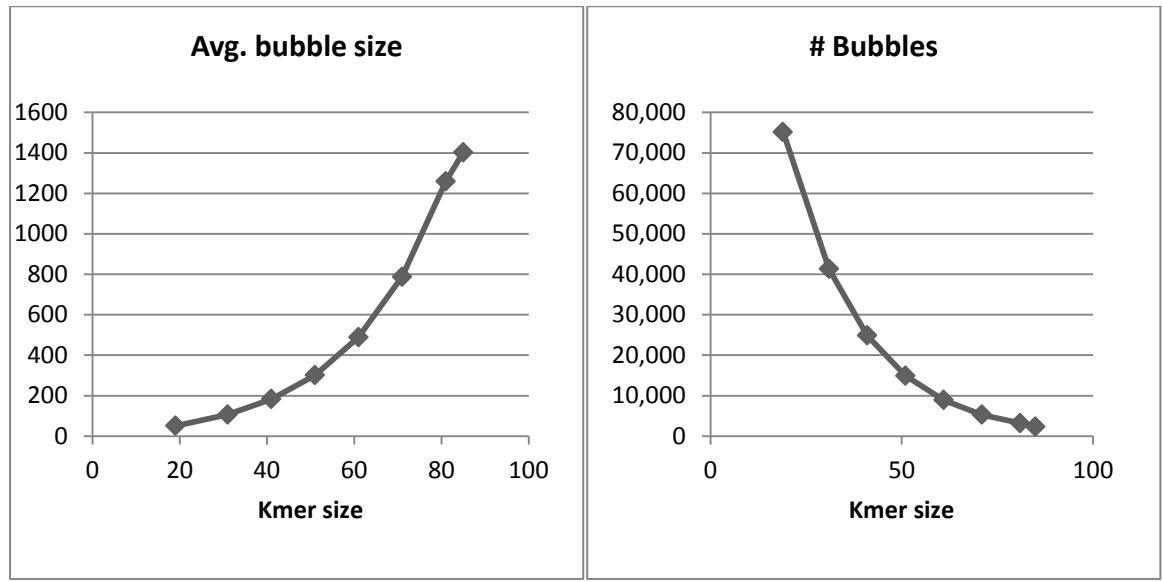

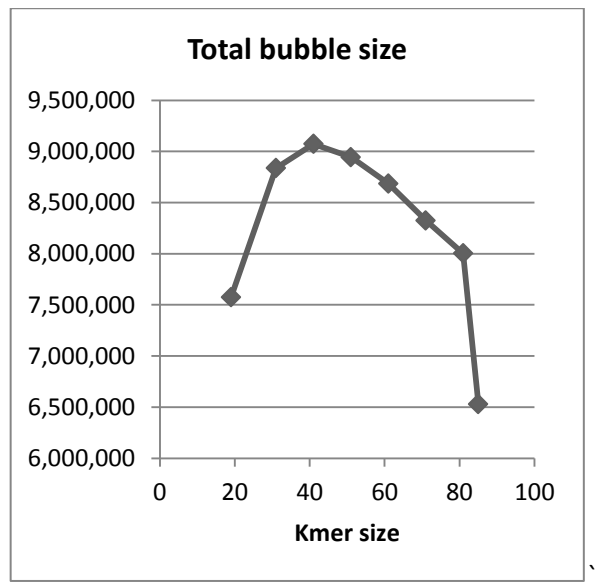

**Fig. 1:** Effect of k-mer size on bubble detection in Meraculous2 from simulated dataset

In Meraculous, when assembling a diploid data set, increasing the k-mer size (*k*) accentuates the effects of heterozygosity.  In a simulated hybrid "pseudo-diploid" *E. coli* dataset where polymorphisms were introduced at random with the probability of 0.05 (*h*), increasing *k* (thus increasing the effective variant k-mer rate *hk*) results in changes in the bubble composition. As seen in plots A and B, increasing *k* results in fewer bubbles and in higher average bubble length (defined as the total length of contigs comprising the bubble). However, as seen in Fig.1c, the *total* length of sequence represented in bubbles decreases sharply as more variant paths fail to fit the bubble definition.

## 2. METHODS

Rather than focusing on isolating bubbles as a way of traversing variant sites, we found that in cases of high polymorphism a more effective and algorithmically simpler solution is to target *non-variant* regions as the source of ambiguity in the graph. In other words, in an ideal extreme case where the haplotypes are so divergent that no common k-mer exists between them, the Meraculous contig graph will resemble that of two *haploid* genomes, each with half the expected sequencing depth. In real life, homozygous regions of the genome will result in collisions in these haploid graphs, but as long as these are relatively infrequent, they can be identified and handled by a small set of heuristics. In particular, we must be able to reliably recognize heterozygous regions as elements in the graph and to distinguish them from homozygous elements. Since in Meraculous2, by design, a pair of variant k-mer extensions will always result in the UUtig termination, contigs will never contain both types of k-mers, *i.e.*, they must be *either* all-variant or all-*non*-variant in their k-mer content. This means that we can rely on relatively stable contig-wide properties like average k-mer depth to help determine the contig's "ploidy." Once this is accomplished, we can construct a haplotype-specific scaffold by requiring that linkages are anchored only on variant-containing segments of the graph.

### 2.1 'Dual Haplopath' approach

In Meraculous-2D we introduce a new optional mode for assembling highly heterozygous genomes, termed here as *Dual Haplopath* mode (or "diploid_mode 2" flag in the program). (Note that genomes with very low heterozygosity will not assemble well using this mode. Instead users should run in *Single Haplopath* mode described below). The key features of this method are:

- Bubbles are identified as in the original Meraculous2 diploid method [Chapman 2016].
- Reads are mapped to bubble contigs only using merBlast algorithm [Chapman 2016].
- Chains of Contig-(Bubble-Contig)$^n$ are constructed as in the original Meraculous2 diploid method [Chapman 2016].
- Exactly two haplotype-specific paths are traced though every Contig-(Bubble-Contig)$^n$ chain using mapped reads, where a read or read-pair mapping to two contigs in different bubbles effectively phase them to a common haplotype.
- From the two paths a pair of contigs, called *p-diplotigs*, is formed and is preserved throughout the assembly process as two representations of a variant region.

- Prior to scaffolding, when re-mapping reads to the p-diplotigs, seeding is allowed only to haplotype-specific areas, *i.e.*, to regions that used to be bubble-contigs, and not to inter-bubble regions. This allows us to maintain haplotype specificity when linking a diplotig to other contigs.
- Since many contigs containing allelic variants are expected to escape the bubble identification algorithm, we categorize the remaining *non*-diplotig contigs as either "full depth" or "half depth" (abbreviated FD and HD). This classification is based on each contig's average k-mer depth relative to the empirically defined diploid depth threshold (established as the minimum between half-depth and full-depth peaks in the overall k-mer frequency distribution), and is relied upon when making scaffolding decisions.
- During the scaffolding stage, we identify FD-HD "link collisions" where an FD-node (a contig or a singleton scaffold) is reciprocally linked to two equidistant conflicting HD-nodes (p-diplotigs are also considered HD here) (Fig. 2). This type of collision is assumed to represent a boundary between homozygous and heterozygous regions and is resolved by creating a "copy-object" and reassigning one of the conflicting links to it. In this manner we introduce non-uniqueness into our contig space in order to maintain a "parallel," two-haplotype assembly, which is a significant deviation from the original Meraculous model in which every contig must consist of strictly unique k-mers. Note that we reject links between nodes where the k-mer depth drops off by an excessive amount, namely by more than the amount equal to the canonical diploid peak depth, as these are likely to be links between single copy and collapsed multi-copy sequences (or other discontinuities in copy number), as this greatly confounds scaffolding and can lead to mis-assembly. Conceivably, it is still possible that a collapsed low-copy repeat representing a naturally duplicated variant allele could be mistaken for a homozygous region and thus could end up getting duplicated, but this should not, in principle, hinder the assembly. On the contrary, expanding a collapsed repeat should reduce linkage ambiguity, granted that sufficient amount of read pairs spanning the repeat unit exist in the data set. (See Fig. 2)

- We don't do any active haplotype phasing inside scaffolds, instead leaving that to happen naturally since with high polymorphism density, variant contigs are expected to connect to contigs of the same haplotype. However, under certain conditions haplotype "cross-overs" can occur throughout the assembly.
- At the end, the final set of scaffolds represents both haplotypes and is thus expected to be roughly twice the size of the individual haplomes. With highly heterozygous genomes, this also has the merit of preserving the diploid variation with maximum contiguity, which can be useful in the downstream analysis.

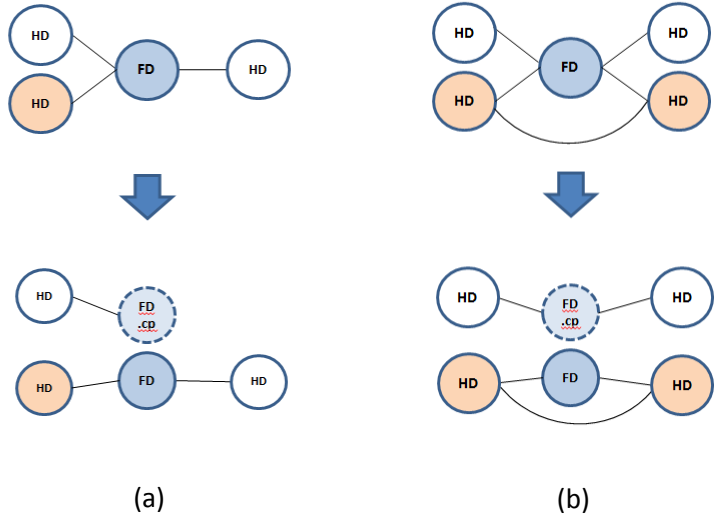

**Fig. 2:** Scaffolding in Dual Haplopath mode. A collision between links to a **full-depth** (FD) contig, originating from **half-depth** (HD) contigs (or p-diplotigs, which are always haplome-specific), signals a boundary between a variant and a homozygous region. In such cases, the corresponding end of the FD contig is marked as a **FD-HD collision** end.

Fig (a): During scaffolding, when evaluating best linkage candidates of an object (orange node), if a candidate *target* end (blue node) is marked as a FD-HD collision, we resolve it by creating a copy of that entire node and re-assigning to it the conflicting links.

Fig (b): In a FD-contig that is "suspended" between two HD-contigs (orange nodes), FD-HD link collisions stemming from *both ends* of the contig can be resolved simultaneously while maintaining haplotype consistency. In fact, contigs with at least one end marked as a FD-HD link collision are suspended whenever possible and later incorporated into the scaffold evoking the copy-based link collision resolution.

2.2 Enhancements to the original 'Single Haplopath' method

The original method of polymorphism resolution in Meraculous2 is still available in a modified form, and is recommended for genomes with low levels of polymorphism $hk < 1$ (referred to as the "diploid_mode 1" flag in the program). This method relies heavily on bubble detection at the contig level and filters out any remaining non-bubble contigs that fall below the diploid k-mer depth cutoff. The focus here, therefore, is on identifying and masking diploid features that lead to scaffolding ambiguities. In the original Meraculous2 [Chapman 2016], fused chains of contig-(bubble-contig)$^n$, i.e., *diplotigs*, represented a single mosaic path through the constituent bubbles, where the choice between every pair of bubble-contigs was independent from other bubbles and the contig chosen was always the one with the higher depth.  These mosaic diplotigs were then allowed to participate in the scaffolding like regular non-variant contigs.  We found this aspect of the method to be less than optimal for reconstructing a diploid genome and have made enhancements to reduce the mosaicism while essentially maintaining the original "single-path" philosophy.   These enhancements make it possible to produce final scaffolds that are more internally consistent with a single haplotype, rather than being a composite mosaic of both haplotypes. Naturally, this also allows reporting the alternative variant contigs in a separate file or have the two sets combined.   The process can now be summarized as follows:

- The steps of bubble identification, mapping of reads to bubble-contigs, and construction of –*p-diplotig* pairs are identical to the *Dual Haplopath* method described above.  The map of clones-to-bubbles-to-diplotigs is generated.   It will be used for haplotype phasing step during scaffolding.
- Contigs that were not associated with any bubble chain, termed *isotigs*, are filtered based on the empirically determined *diploid depth cutoff*.  The aim here is to eliminate uncaught variants from the scaffolding process, thus simplifying the implied assembly graph. (As stated earlier, in low-polymorphism data sets, these are expected to be mostly small in size and thus most gaps caused by removing them should be easily closable by k-mer "walking" during the gap closure stage [Chapman 2016].)
- When re-mapping reads back to the contigs which passed the diploid depth filter, here, unlike the *Single Haplopath* method, we don't attempt to enforce strictly variant-specific mapping, and allow seeds to occur throughout the contig length.
- Only one contig of each variant p-diplotig pair is used in scaffolding (labeled _P1) while the alternative variant (labeled _P2) is left as a singleton.  Even though read mapping information is generated for each p-diplotig separately, it is used in an aggregate fashion when establishing contig linkage, meaning, the bubble is linked as a whole, but the linkage data is combined and applied to one variant contig only.  This feature is what in effect simplifies the assembly graph and leads to greater long-range contiguity.   In other words, bubbles are again collapsed, but here, unlike Meraculous2, the alternative variant information is not lost.  Keeping the alternative variant p-diplotig around as a singleton allows us to swap it in later, when phasing the variants to a common haplotype  (see Fig 3).
- Once the final round of scaffolding has completed, we re-evaluate scaffolds and their constituent p-diplotigs in terms of their haplotype consistency.  Here we focus on strictly variant-specific read mapping (*i.e.*, reads mapping to individual bubble-contigs).   For every _P1 p-diplotig in a scaffold we check whether sufficient linkage exists anchoring it to the p-diplotig(s) upstream or whether its alternative _P2 variant provides better linkage.  If it's the latter, the _P1 p-diplotig is swapped for the corresponding _P2 variant.  Since _P2 variants are singletons, the swapping doesn't affect any other scaffold (see Fig 4).
- Gap closing proceeds normally using all mapped reads that are thought to extend or project into gaps.
- In the final reported assembly the singleton alternative variants are removed but their list is provided separately, as well as the unfiltered, all-inclusive set of scaffolds.

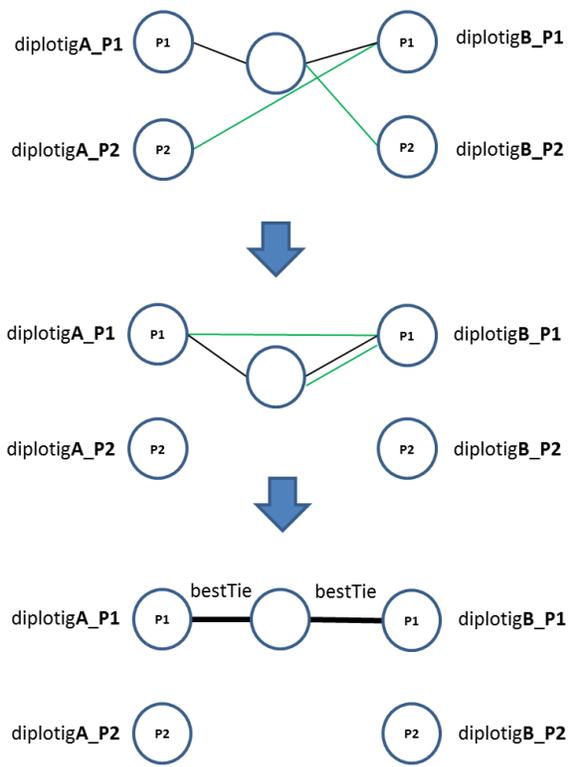

**Fig. 3:** Scaffolding in Single Haplopath mode

Diplotig pairs represent alternative contiguous haplotype variant regions. To construct a single path through a series of nodes involving links between different p-diplotigs, all links to/from _P2 p-diplotigs are transferred to the corresponding _P1 "sister". This in effect "flattens" the graph, allowing determining mutual "best ties" which form the basis for building scaffolds. Existing links to/from _P1 p-diplotigs are preserved, and if a transfer of a _P2 link in this manner results in a significant disagreement in the distance estimations, a mutual best tie cannot be formed. This helps distinguish repeat units from true haplotype variants.

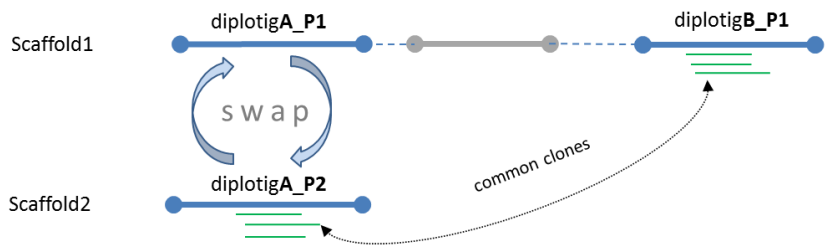

**Fig. 4:** Haplotype phasing in Single Haplopath mode.

2.3 Validation of the Dual Haplopath method

Evaluating assembly quality is often as challenging as the assembly problem itself, in part because quality of a fragmented and incomplete reconstruction of a new genome is hard to define in a small number of clear terms [Earl 2011, Bradnam 2013]. Contiguity metrics like the N50/L50 are not in themselves strong measures of assembly quality since contigs and scaffolds can and do contain mis-joins, which can artificially inflate metrics. Comparing an assembly to a validated reference sequence is a luxury rarely available when dealing with highly heterozygous diploid genomes as there are very few published assemblies of these types of genomes that are themselves sufficiently complete and accurate. However, we can talk about the "usability" of an assembly in terms of how well it allows a downstream user to extract biologically relevant information from it.

In this respect, we found it helpful to use "pseudo-diploid" data sets where the source reads come from two haploid individuals (either real or simulated) with a well-defined nucleotide divergence and the homologous genomes are divergent enough to imitate highly heterozygous haplomes of a hypothetical diploid genome. This allows us to achieve several goals. First of all, we can evaluate the diploid assemblies' fidelity or "purity" with respect to the two known individual "progenitor" assemblies, with the expectation that the latter is typically a lot more contiguous and complete. Second, mis-assemblies are easier to detect and evaluate against a high quality *haploid* assembly, and several assembly assessment tools exist that work with haploid references. (We found no available tools for comparing dual-haplotype assemblies.) Third, we can assess each assembly's practical usefulness in terms of the number of unique functional predictions that could be made using JGI's annotation pipelines and to see how well that approaches the corresponding annotation of the individual haploid progenitors.

To benchmark our results, we chose two recently published diploid-aware assemblers, dipSPAdes (v.3.9) and Platanus (v.1.2.4). For data sets with known and assembled haploid or haploid-like progenitors we measured assembly "purity" using two approaches, one alignment-based and the other based on shared k-mer abundance. The alignment-based method used QUAST v.4.2 with identity cutoffs set at 95, 99, and 100 percent to infer what fraction of the assembly as a whole has high identity with one of the progenitors, irrespective of contiguity. The k-mer based method, on the other hand, used *kcompress* and *bbduk* (both part of the BBtools package: http://jgi.doe.gov/data-and-tools/bbtools/) to measure the "pure" contiguity of contigs/scaffolds by requiring that an assembled unit consist entirely of k-mers shared with a given progenitor. Since the likelihood of haplotype crossover tends to increase with scaffold size, we normalized the length by cutting the scaffolds into adjacent 1 kb fragments and measuring the k-mer purity in these fragments. This way, the effect of a crossover event on the purity metrics is independent of the particular assembler's affinity to longer contigs/scaffolds.
The results are summarized below.

2.3.1 Pseudo-diploid *Escherichia coli*

For our simplest model, we used the well-studied *E.coli K12* genome (acc: NC_000913.2 gi: 49175990) and created "pseudo-haplomes" by introducing random single-bp variation at rates 0.01,0.02,0.05, and 0.1 into the sequence. Among the variant sites introduced, 80% were 1-5 bp substitutions and the remaining 20% were single-bp. indels. Illumina-like paired reads were simulated from both "haplomes" with a realistic sequencing error distribution and three Gaussian populations of insert sizes (+/- standard deviation) at 225+/-25, 450+/-50, and 4000+/-50 bp. The data sets were assembled with Meraculous-2D using the k-mer size of 75 bp and the two diploid assembly modes. Three rounds of scaffolding (one with each insert size) were performed to produce the final scaffold N50 values appearing in figure 5. The assemblies were confirmed to be free of major mis-assembly by mapping the scaffolds back to the original *pseudo*-haplotypes and visually inspecting the map. As seen in Figure 5, the *Dual Haplopath* approach produced a great increase in the scaffold N50 when the SNP rate exceeded 1/k, which reflects the mode's ability to circumvent the bubble detection limitation posed by the high SNP rate. Note that since variants are introduced in an uncorrelated fashion, this synthetic dataset is not necessarily a good model for the diploid genome of a sexually reproducing organism, since the distribution of heterozygous positions in an interval is geometric rather than normal [Rosenberg 2002].

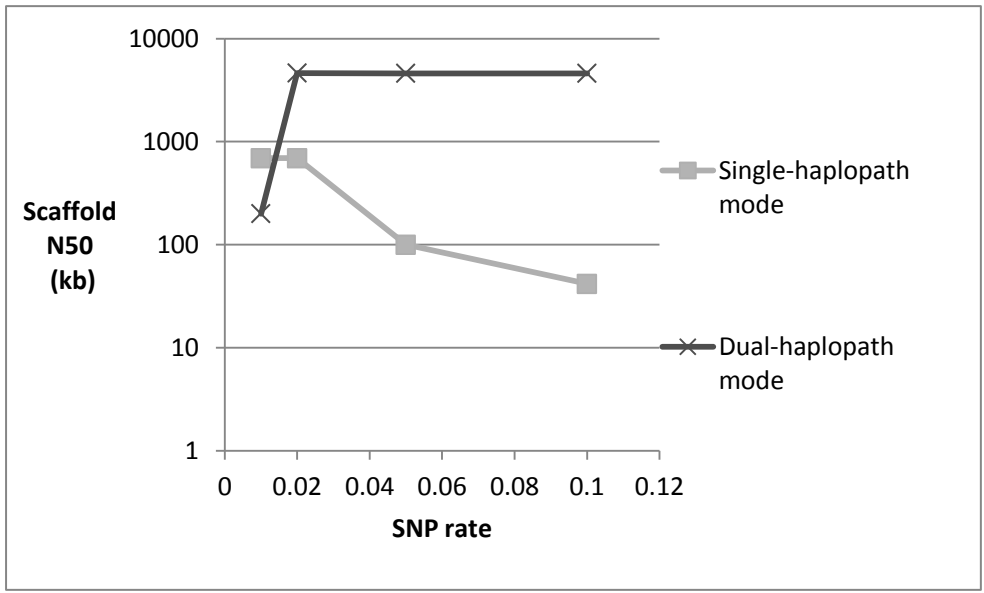

**Fig. 5:** Assembly of synthetic diploid E.coli datasets using the two diploid modes in Meraculous-2D at fixed k=75.

2.3.2. Pseudo-diploid *Schizophyllum commune*

*Schizophyllum commune* is a haploid fungus, but any two individuals are estimated to diverge by ~7% {Safonova 2015}. A number of *S. commune* strains have been sequenced and deposited in NCBI's Sequence Read Archive (SRA), which provides a good starting point for creating a pseudo-diploid data set where a pair of strains sequenced at similar depth can be combined and assembled de-novo as a diploid. We obtained the sequence data from S.commune strains A8 (SRR1548681) and B3 (SRR1548670), which is the same data set that was used in the original dipSPAdes publication {Safonova 2015}. While Meraculous2D produced an assembly of lower N50 compared to dipSPAdes (see Fig.6), the purity metrics in Figure 7 show much higher progenitor fidelity compared to either dipSPAdes or Platanus. When Platanus was run with a more aggressive set of parameters (-u 0.7 –s 1), the N50 exceeded that of our assembly, but haplotype purity suffered significantly.

Automatic gene calling and annotation of the Meraculous2D assembly using JGI's fungal annotation pipeline (http://genome.jgi.doe.gov/programs/fungi/FungalGenomeAnnotationSOP.pdf) detected very similar numbers of unique functions compared to both dipSPAdes and Platanus (results not shown here), which suggests that the lower overall N50 length did not hinder this important aspect of the downstream analysis.

Potential mis-assemblies were detected using QUAST v.4.2 with the following parameters: -s --fragmented -a all --ambiguity-score 0.95 --min-identity 95. Since dipSPAdes and Platanus aim to reconstruct any given variant region as a single sequence, to reduce the chance of mis-mapping we used a single haploid assembly of *S. commune* A8 as the reference, while for Meraculous2D assemblies we combined A8 and B3 references. This approach resulted in the highest aligned fraction and the lowest mis-assembly rates reported by QUAST across all three test cases.

| Genome | Assembly Method | Scaffold N50 (kb) | Contig N50 (kb) | Total size in scaffolds (kb) | Misassemblies (QUAST) |
|---|---|---|---|---|---|
| ***E.coli*-0.01** (1% pseudo-diploid) | Meraculous2D – diploid_mode 1 | **692** | **133** | 4,604 | 0 |
| | Meraculous2D – diploid_mode 2 | 420 | 88 | 9,267 | 0 |
| ***E.coli*-0.02** (2% pseudo-diploid.) | Meraculous2D – diploid_mode 1 | 692 | 59 | 4,590 | 1 |
| | Meraculous2D – diploid_mode 2 | **4,639** | **344** | 9,245 | **0** |
| ***E.coli*-0.05** (5% pseudo-diploid.) | Meraculous2D – diploid_mode 1 | 111 | 17 | 4,265 | 0 |
| | Meraculous2D – diploid_mode 2 | **4,640** | **4,640** | 9,251 | 0 |
| ***S.commune*** (~7% pseudo-diploid) | dipSPAdes | **85** | **85** | 40,245 | 614 |
| | Platanus (defaults) | 9 | 8 | 47.156 | 76 |
| | Platanus (aggressive) | 35 | 26 | 41,094 | 172 |
| | Meraculous2D – diploid_mode 2 | 18 | 17 | 71,000 | **30** |

**Fig. 6:** Assembly size and contiguity comparison.

| Genome | Assembly Method | Sum of "pure" fragments (kb) | Assembly Fraction Aligning to Haploid References (A8+B3 assemblies combined) | | |
|---|---|---|---|---|---|
| | | | Id. 100% | Id. 99% | Id. 95% |
| **S.commune** (~7% pseudo-diploid) | dipSPAdes | 15,887 | 0.16 | 0.23 | 0.98 |
| | Platanus (defaults) | 32,375 | 0.47 | 0.82 | 0.96 |
| | Platanus (aggressive) | 24,459 | 0.18 | 0.53 | 0.97 |
| | **Meraculous2D (diploid_mode 2)** | **66,570** | **0.96** | **1.0** | **1.0** |

**Fig. 7:** "Purity" evaluation of *S.commune* assemblies.

For each assembly, the final scaffolds were divided into windows of 1000 bp, and each window's kmer content was compared against that of the known haploid progenitor strains A8 and B3. A window was considered "pure" if <u>all</u> of its constituent k-mers had a matching k-mer in a single progenitor assembly. The total sum of such windows is thus a measure of how faithfully the individual haplotypes are represented in a given assembly. The total length of final scaffold sequence that aligns with 100% identity to the *combined* A8+B3 progenitor reference is also provided as another measure of assembly fidelity that is not dependent on scaffold N50 size.

## 3. DISCUSSION

The two modes of Meraculous2D introduced here offer a choice of approaching the diploid assembly problem in complementary ways. In the *Single Haplopath* method, the underlying expectation is that variants are isolated and most bubbles are caught and are represented in annotated contig (p-diplotigs) pairs. The emphasis is therefore strongly on emulating haploid scaffolding by "flattening" variant-induced linkage conflicts. As we demonstrated here, when the rate of variant k-mers is large, much of the variation is not being captured, which greatly hinders subsequent scaffolding. The *Dual Haplopath* approach addresses this by treating the contigs as fragments of two distinct haplomes. The emphasis then becomes (1) identifying *non-variant* (homozygous) regions that "collapse" in the local assembly graph, and (2) resolving such collapses. If this can be done with high fidelity, the assembly task becomes similar to assembling two haploid genomes together, and the output is expected to be roughly twice the size of the diploid genome. If a haploid-like reference assembly is desired where the final sequence is non-redundant in allelic composition, there exist a number of post-assembly tools that generally rely on alignment-based approaches to achieve this goal (e.g., HaploMerger{Huang et al 2012}; see also {Kim et al. 2007} and {Small et al. 2007}). In this respect, Meraculous2D offers the benefit of preserving all variants so that the allelic relationships could be better reconstructed by downstream analyses.

Since variant frequency can vary across a genome, different genomic regions may be more suitable for one method or the other, so it often makes sense for a user to try both modes. If there is evidence that the variation is highly uneven (*e.g.*, the bubble-contig size distribution is very broad), one may want to first perform an assembly in *Single Haplopath* mode, and then identify and remove from the k-mer set all k-mers that assembled into long scaffolds, along with those from the corresponding alternative variant singletons (running merCounterTh standalone on the contigs will help accomplish this). In this manner, a reduced set of k-mers is produced that is likely to derive from highly heterozygous regions. After that, using the `bootstrap_run.sh` script and the –restart feature of `meraculous.pl`, the user can begin another assembly starting with the reduced k-mer set and using the *Dual Haplopath* mode to tackle the regions of high polymorphism. Since no common k-mers exist between the two assemblies, the results can be safely merged.